\documentstyle[eqsecnum,aps]{revtex}

\begin{document}
\draft
\preprint{PUC-Rio, UFRJ}
\twocolumn[\hsize\textwidth\columnwidth\hsize\csname
@twocolumnfalse\endcsname
\title{  New Magneto-Roton Modes in the "Ribault Phase" of the\\ Ultra Quantum Crystal }
 \author{Pascal Lederer }
\address{Depto de Fisica, PUC-Rio, , and Instituto de Fisica, Universidade Federal do Rio de Janeiro, Ilha do Fundao, Rio de Janeiro\\On leave from
Physique des Solides,U. P. S., F91405 Orsay, France( Laboratoire associ\'e au CNRS)}
\date{Aug. 8, 1997}
\maketitle
\begin{abstract}
 The Ultra Quantum Crystal phases observed in quasi-one-dimensional conductors of the Bechgaard salts  family under magnetic field exhibit both Spin Density Wave order and a Quantized Hall Effect. As a result, they also possess a Magneto-Roton mode within the single particle gap, and the usual Goldstone modes. The sign reversals of the Quantum Hall Effect have recently been explained within the Quantized Nesting Model. I show here that the sign-reversed phases of the Ultra Quantum Crystal (the "Ribault Phases") have novel properties compared to the majority sign  phases. They exhibit at least two Magneto-Roton modes, the energy of which, relative to the single particle gap, vary linearly with the magnetic field, with opposite signs, in contrast to the field dependence in the majority sign phases.                                                                         
\end{abstract}
\vskip2pc]
\section{Introduction}

Organic conductors of the Bechgaard salts family, $(TMTSF)_2 X$ where $TMTSF$ =
tetramethylselenafulvalene are  quasi-one-dimensional (quasi-1D) systems, which have been found over the last few years to exhibit fascinating properties under magnetic field \cite{review,gorkov,gm91,pl96,hlm1}. The typical hierarchy of their transfer integrals is: $t_a=3000K$,  $t_b=300K$, $t_c=10K$
In three members of this family ($X=ClO_4,PF_6,ReO_4$), the metallic phase is destroyed by a moderate magnetic field $H$ applied along the $c$ direction, perpendicular to the most conducting planes ($a,b$). A cascade of  magnetic phases, separated by first order transitions appear as the field intensity is stepped up: within each sub-phase, which I have called "Ultra Quantum Crystal" (UQC in the following) a Spin-Density Wave  Phase ( FISDW) (i.e. a Quantum Crystal)
is stabilized  with a peculiar electronic structure, characterized by a small number of exactly filled Landau levels (bands in fact) \cite{pl96}. Each UQC sub-phase exhibits a Quantized Hall conductivity, which is the first example of a Quantum Hall Effect in a 3D system.
This cascade of quantized phases result from an interplay between the nesting properties of the Fermi Surface (FS), and the quantization of 
electronic orbits in the field: the wave vector of the SDW varies with field so that unpaired carriers in a subphase are always organized in
 completed filled Landau bands. As a result the number of carriers in each subphase is quantized, and so is the Hall conductivity: $\sigma_{xy}=2Ne^2/\hbar$. The factor 2 accounts for spin
degeneracy \cite{review} \cite{pl96}. The condensation of the UQC phases results from the peculiar electronic strucure of open Fermi Surface metal under magnetic field: because of Lorentz force, the electronic motion becomes periodic  and confined along the high conductivity direction of the chains ({\bf a }direction)\cite{gorkov}.
 The periodic motion of the electrons in real space
 is characterized by a wave vector $G=eHb/\hbar$, $b$ being the interchain distance. (In the rest of this Letter, wave vectors will be expressed in units of $G$). As a result, the static 
bare susceptibility of the normal 
phase, $\chi _0({\bf Q})$ can be expressed 
as a sum over weighted
 strictly 1D bare susceptibilities which diverge at quantized
  values of the longitudinal component of the wave vector $Q^n_{\vert  \vert}=2k_F+n$\cite{gm91,pl96,hlm1}.
 The largest divergence signals the appearance of a SDW phase with quantized vector $Q_{\vert  \vert}=2k_F+N$. This  Quantized Nesting Model (QNM)  \cite{hlm1} describes most of the features of the phase diagram in a magnetic field. It
 has been shown recently to explain  the experimental
 observation of  the Hall plateaux sign reversal when the field varies\cite{zm}.
 Most plateaux exhibit the same sign. (By convention I will refer to these plateaux
 as positive ones).  The sign reversal has been discovered by Ribault in $(TMTSF)_2ClO_4$
 under certain conditions of cooling rate\cite{rib2}. Negative plateaux
 have been reproduced and also found in $(TMTSF)_2PF_6$ where their existence
 depends crucially on pressure \cite{pivetau,cooper,bali}. Recently,
 Balicas et al \cite{bali} have shown that there exists a range of pressure 
 for which, in the $PF_6$ salt, the sequence of observed plateaux when the field
 decreases can be identified with the quantum numbers $N=1,2,-2,3,4,5,6,7$.
 A more ancient experiment has shown a sequence of phases $N=1,2,-2,4,-4,5,6$\cite{cooper}. Hereafter, I will refer to the UQC Phase with negative Hall number as "Ribault Phases".

 Zanchi and Montambaux\cite{zm} have shown that the negative plateaux can be understood within the QNM assuming the dispersion relation in the normal phase to be:

\begin{eqnarray} \label{model}
\epsilon({\bf k})& =& v_F(\vert k_x\vert - k_F)+ \epsilon_{\perp }({\bf k_{\perp}}),\\
\epsilon_{\perp }({\bf k_{\perp}})& =& -2t_b\cos k_yb  -2t_c\cos k_zc -
2 t'_b\cos 2k_yb \nonumber \\ &&-2t_3\cos3kyb -2t_4\cos4kyb
 \nonumber
\end{eqnarray}

$\epsilon(k_{\perp})$is a  periodic function which describes a warped FS.
With $t_3=t_4=0$,  Eq.(\ref{model} ) cannot lead to sign reversals, as $sign(N)=sign(Q_{\vert  \vert}-2k_F)=sign(t'_b)$ \cite{zm}. However small values of $t_3\simeq .2t'_b=2$K, and $t_4=.2$K are sufficient to account for the experimental results of Balicas et al \cite{bali} \cite{zm}. 

The  normal metal-FISDW instability line  $T_{cN}(H)$ is given by: 
\begin{equation} \label{ki}
\chi _0({\bf Q},T_{cN}, H)=
\Sigma _nI_n^2 (Q_{\perp })\chi _0^{1D}(Q_{\vert  \vert}-n, T_{cN})=1/\lambda
\end{equation}
$\lambda$ is the electronic interaction constant. Eq.(\ref{ki})
 exhibits the structure of $\chi _0$ as the sum of one
 dimensional terms $\chi _0^{1D}$ shifted by the magnetic field wave
 vector $G=eHb/\hbar$. 
$\chi_0^{1D}\propto -\ln(\max \{ v_F(2k_F-q),T   \}/\epsilon_F)$. 
In Eq. (\ref{ki}),  the coefficient $I_n$ depends on the dispersion  relation and H:
\begin{eqnarray}
I_n(Q_{\perp})   &= &\nonumber  \\ 
<\exp i\left[(T_{\perp}(p+Q_{\perp}/2)+ 
  T_{\perp}(p-Q_{\perp}/2))+np
 \right]> &&
\end{eqnarray}
where $T_{\perp}(p) = (1/\hbar \omega _c)\int_0^p\epsilon_{\perp}( p')dp'$
and $<...>$ denotes the average over p.

Let me define a generalized  instability temperature $T_{N \pm  m}$:
\begin{eqnarray} \label{Tc}
1/\lambda &=& I^2_{N\pm m}(Q^N_{\perp } \pm q_{\perp }) \ln \left( \frac{2\gamma E_0}{\pi T_{N \pm m}}  \right) +  \nonumber \\ &&                      
\sum_{n \neq  0}  I^2_{N\pm m +n} (Q_{\perp }^N \pm q_{\perp})\ln \left(    \frac{E_0}{\vert n\vert \omega _c}\right)
\end{eqnarray}
In Eq.(\ref{Tc}), $\gamma$ is Euler's constant, $E_0$ is a high energy cut-off, and $\omega_c =v_FG/2$.
For $m=0$ and $q_{\perp}=0$, $T_{N \pm m}=T_{cN}$, the ordering temperature for the $N$th subphase. For $m\neq 0$, $T_{N \pm m}(q_{\perp})$ generalizes the definition of the critical temperatures on either side of phase $N$ in the $(T,H)$ plane. $T_{N \pm m}(q_{\perp})$ are at most equal to the virtual transition lines $T_{ N\pm m}$ which can be drawn in the $N$th subphase part of the phase diagram and which represent virtual transition lines to phases with slightly larger free energy than the $N$th subphase\cite{pl87}. In the ($T,H$) plane, there is an infinite number of continuous lines  crossing the phase diagram. The upper limit of this family is the actual (continuous non analytic) transition line from the normal metal to the UQC; this  line coincides piecewise with the transition lines  labelled by  the successive integers describing the Quantum Hall conductivity. See Fig.(1) . An example of computed network of transition lines was given in \cite{lm}

Using this network of real and virtual transition lines, I  showed, with Poilblanc, that  the UQC collective modes exhibit, aside from the usual Golstone modes linear in wave vector, at least one Magneto-Roton (hereafter MR) mode within the single particle gap, located at $q_{\vert  \vert}=G=1$, $q_z=\pi /c$, and $q_y$ some optimum $q_y$\cite{pl87,dppl}. This mode, together with the usual Goldstone modes, is the signature of the novel nature of the electron-hole condensate in UQC: a Spin Density Wave driven by orbital quantization and an Integer Quantum Hall system driven by electronic interactions. The MR mode is both a consequence of  quasi 1D electronic periodic orbital motion, gauge invariance, and the long range crystalline order of the Field Induced Phase\cite{dppl}.

One obvious question comes to mind: {\it are the Ribault phases the same objects as the phases          with positive plateaux?} At first sight, the answer seems to be positive: both have  a periodic modulation of Spin Density,  both exhibit Quantized Hall Plateaux, the only difference being the sign of the carriers, in turn connected to a quantized wave vector $Q_{\vert  \vert}$ which is larger (smaller) than $2k_F$, for N positive (negative).

The purpose of this Letter is to point out that {\it the Ribault Phase of the UQC differs from the usual UQC in  one important aspect: its collective modes exhibit at least {\bf two}, (sometimes {\bf four},or more), low lying MR modes within the gap}, with different wave vector components  $q_{1\vert \vert} =M_1$ and $q_{2\vert \vert}=M_2$ at the magneto-roton  minima with $M_1,M_2$ integers $>1$, (sometimes $q_{i\vert \vert} =M_i >1 $, $i$ from 1 to 4,etc.,) as opposed to one low-lying mode for the usual UQC, with wave vector
$q_{\vert \vert}=1$. Both MR minima, relative to the single particle gap, {\it vary with field with opposite sign of the derivative versus field, as opposed to an almost field independent MR mode in the usual UQC}.
To my knowledge this is the first example of such a rich collective mode structure in a quantum condensate. This will result in possibly markedly different physical properties for the Ribault Phase , as well as its neighbouring phases
as compared with the usual UQC.\cite{pl97,plcmc}

In order to prove my point, let me recall a few results on the UQC collective mode theory\cite{pl87,dppl}
\section{New MR Collective Modes}

 The   collective modes    are obtained  from the poles
 of the spin-spin correlation function in the ordered phase,  in the RPA \cite{pl87,dppl}. 
The equation  is:

\begin{eqnarray} \label{rot}
&(1-\lambda \hat{\chi }^0_{+-}({\bf Q_N +q}, \omega ) )(1-\lambda \hat{\chi }^0_{+-}({\bf  Q_N-q}, \omega )) & \nonumber \\ &         -      \lambda^2 \hat{ \Gamma}^0_{+-}({\bf q}, \omega )\hat{ \Gamma}^0_{-+}({\bf q},\omega ) =  0&
\end{eqnarray}
with ${\bf q= Q - Q_N}=$ collective mode wave vector. In Eq.(\ref{rot})
  $\hat{\chi }_{+-}^0$ are the irreducible bubbles renormalized by all
 possible scatterings on the mean field potentials connected to the
 various gaps.
Likewise
 $\hat{\Gamma}^0_{+-}({\bf q}, \omega )$
is the extraordinary bubble, also renormalized with all possible
 scatterings. 

The simplest approximation resums to all orders the gap $\delta_N= \Delta I_N$ at the Fermi level and takes all other gaps into account to second order in perturbation\cite{pl87,dppl}. Then
\begin{eqnarray}
\hat{\chi}^0_{+-}({\bf Q}_N+ {\bf q}, \omega)&=&\Sigma_n I^2_{N+n}(Q_{\perp }^N+q_{\perp }) \tilde{\bar{\chi }}^0\left( n -q_{\vert  \vert}, \omega \right)
\end{eqnarray}
and
\begin{eqnarray}
\hat{\Gamma}^0_{+-} ({\bf q}, \omega )=& &\nonumber \\    \Sigma_n I_{N+n}(Q_{\perp}^N +q_{\perp})I_{N-n}(Q_{\perp }-q_{\perp })  \tilde{\bar{ \Gamma^0}} \left(   n - q_{\vert  \vert}, \omega   \right) &&
\end{eqnarray}
$\tilde{\bar{\chi}}^0$ and $\tilde{\bar{\Gamma}}^0$ are for $n=0$ the
 objects discussed in \cite{lra} in connections with collective modes of SDW. For $q\ll 1$ 
 and $\omega \ll 2\delta_N$ Eq.(\ref{rot}) describes the usual 
phase and amplitude modes $\omega^2=v_F^2{\bf q}^2$ and 
$\omega^2=v_F^2{\bf q}^2 + 4\delta^2_N$.
  {\it New physics appears for
$q_{\vert  \vert} =m +\delta $}, with $\delta \ll 1$ and $m$ integer. In
that case, $\hat{\chi}^0_{+-}({\bf Q}_N+ {\bf q}, \omega )\neq
\hat{\chi}^0_{+-}({\bf Q}_N- {\bf q}, \omega )$, so that {\it Eq.(\ref{rot})
 does not factorize anymore.}

Then an interaction with the gap at $N\pm m$  allows the collective mode
to propagate in a medium almost identical to the case $m=0$ and
$q_{\vert  \vert}\ll 1 $. A second interaction allows the outgoing oscillation to
retrieve the momentum lost with the first interaction. {\it The mode
with $m\neq 0$ would have exactly the same energy as that with $m=0$ and
$q_{\vert  \vert} \ll 1$ if all $I_N$ were equal}. Such is not the case, so that
the {\it phase and amplitude modes of the order parameter are not
decoupled anymore} for $m\neq 0$ and, {\it  instead of a zero energy mode at
$q_{\vert  \vert}=m$, a local minimum appears}. More precisely Eq.( \ref{rot}) reduces to:

\begin{eqnarray} \label{rot2}
\left( \ln \left(  \frac{2\gamma E_0}{\pi T_{N+m }} \right) -   \tilde{\bar{\chi_0}}(\delta,\omega)                                             \right)&& \nonumber  \\  \times
 \left(  \ln \left(  \frac{2\gamma E_0}{\pi T_{N-m}} \right) -\tilde{\bar{\chi_0}}(\delta, \omega)   \right) & =& 
 \left( \tilde{\bar {\Gamma}} (\delta ,\omega )    \right)^2 
\end{eqnarray}

where $T_{N \pm m}$ is defined in Eq.( \ref{Tc}), 
$ q_{\vert  \vert}=2k_F +m +\delta$,                                            and $\delta \ll 1$.
    For simplicity, I restrict the discussion here to T=0K\cite{pl97},\cite{plcmc}. Then \ref{rot2} yields, setting $x=\omega_{MR}(m,\delta=0)/2\delta_N $, ($x<1$) \cite{pl87}:
\begin{eqnarray} \label{rot3}
\left(\ln \left(\frac{T_{cN}}{T_{N+m}}\right)   -(x^2-1/2)h(x) \right) && \nonumber  \\ \times
 \left( \ln\left(  \frac{T_{cN}}{T_{N-m}}\right)-(x^2-1/2)h(x) \right) =h^2(x)/4     &&
\end{eqnarray}
 where $h(x)=\frac{sin^{-1}x}{x(1-x^2)^{1/2}}$.
Using (\ref{rot2}) and (\ref{rot3}), one proves the existence of at least 
one low lying MR mode at $m=1$ in the usual UQC case with no Ribault
 phase \cite{pl87}. The possibility of other MR minima  with $m>1$ was
mentionned, but until now no proof was given for their existence at
 energies well inside the single particle gap\cite{pl87}. The field 
dependence of the MR mode at $m=1$ is easy to find when $\epsilon_{\pm 1}=(T_{cN}-T_{N\pm 1})/T_{cN}\ll 1$. The MR
 minimum is : $x^2_0=(\epsilon_{+1}+\epsilon_{-1})/2$. Since $\epsilon_{+1}$
 and $\epsilon_{-1}$ have opposite and nearly equal variation with field 
within the  phase N (see \cite{lm}), $x^2_0$  is almost constant 
 within a given phase, equal to the value of $\epsilon_{\pm 1}$ at their crossing point .

Consider now the Ribault Phase  $N=-2$ studied by Balicas et al. \cite{bali},  with the sequence of quantum numbers 1,2,-2,3,4,5,6,7. The lowest lying modes are not at $q_{\vert  \vert}=1$ any more, but at
$q_{\vert  \vert,\alpha}=4$ and $q_{\vert  \vert,\beta}=5$.  Consider 
 one of these, say $q_{\vert  \vert,\alpha}$. Define $\epsilon_{\alpha } (H)=(T_{c,-2}-T_2)/T_{c,-2}$ for $T_{c,-2} \geq T_2$ ; $\epsilon_{\alpha}$ goes to zero  at the transition between phases $N=-2$ and $N=2$; it varies roughly linearly with field, with a negative slope of order a few tenth of K per T; assume for simplicity $\epsilon_{\alpha}\ll 1$ in the whole Ribault phase. Define also $L_{\alpha }=\ln(T_{c,-2}/T_{-6})$. $L_{\alpha}$ is certainly larger than 1 and slowly varying in the whole Ribault phase. Assume it is constant, for simplicity, with no loss in generality. The equation for the corresponding MR minimum $a(H)=\omega_{MR}(4) /2\delta_{-2}$ is:

\begin{eqnarray}\label{rot4}
[\epsilon_{\alpha}-(a^2-1/2)h(a)][L_{\alpha}-(a^2-1/2)h(a)]&=&h^2(a)/4
\end{eqnarray}

In the (realistic) limit $L_{\alpha}\gg 1$, there is always a solution $a=a_0 +s_{\alpha }\epsilon_{\alpha}(H)$, where $a_0\leq 1/\sqrt2$ and $s_{\alpha}\simeq 1$ . At most,  $a$ may have a weak additionnal field dependence $\propto L^{-1}_{\alpha}$.
Similarly, for $q_{\vert  \vert,\beta}$,  I find a second MR minimum $b=b_0 +s_{\beta}\epsilon_{\beta}(H)$, with $\epsilon_{\beta}=(T_{c,-2}- T_3)/T_{c,-2}$, which has the same magnitude but a slope opposite to that of $\epsilon_{\alpha}$. See Fig. (1). Now $L_{\beta}=\ln(T_{c,-2}/T_{-7})$ and is  large. Around each  minimum, the MR dispersion is strongly anisotropic \cite{pl97}. For $\delta \ll 1$, $\omega_{MR}^2(m,\delta)=\omega_{MR}^2(m,0)+ s(v_F\delta)^2$,($s\sim 1$)\cite{pl87}.
Interesting things should happen when the two magneto-roton minima , which have commensurate parallel components of their wave vectors (here 4 and 5) cross, as they are likely to do, somewhere within the Ribault phase\cite{pl97}. Fig.(2) summarizes the results for the Ribault phase
$N=-2$.

The Ribault Phase contaminates the neighbouring normal phases: within the $N=3$ phase, close to the (3,-2) transition, Eq.(\ref{rot3}) , on top of the mode at $q_{\vert  \vert}=1$, yields a MR mode at $q_{\vert  \vert}=5$. Since $\ln(T_{c,3}/T_8 )$ is not small \cite{lm,plcmc}, the field dependence of that mode should be similar in magnitude to the corresponding mode in the Ribault phase, but opposite in sign. The mode at $q_{\vert  \vert}=1$ should keep its much weaker field dependence, because the line $T_{2}(H)$  passes through the $N=3$ phase rather close to the actual metal-UQC instability line.
The Ribault phase also exerts its influence on the normal UQC when, although thermodynamically unstable, it is close to being stable. This situation 
 can be experimentally realized by tuning the pressure. The $N=-2$ phase is actually quite close to being stable even when $t_3=t_4=0$ in Eq.(\ref{model}). (See ref. \cite{lm}). As the virtual transition line to a Ribault Phase , say with $N=-2$ nears the usual UQC-normal metal transition lines, say $N=3$ and $N=2$ from below, secondary MR modes appear from the bottom of the conduction band, at the wave vector of the relevant transition: within the $3$  (resp. $2$) phase, at $q_{\vert  \vert}=5$, (resp. $q_{\vert  \vert}=4$). Call $\eta_3$ (resp. $\eta_2$) the smallest relative distance between the virtual line $T_{-2}$ and the lines $T_{c3}$  (resp $T_{c2}$). Then the lowest new MR mode energy is: $  \omega_3(\eta_{3}, 5 )/2\delta_3= \omega_3(0,5) /2\delta_3+ s_{\eta_{3}} \eta_3$
 (resp. $\omega_2 (\eta_2, 4 )/2\delta_2= \omega_2(0,4)/2\delta_2 + s_{\eta_2}\eta_2$). 
 The field dependence follows along similar lines.

 I expect a richer structure yet of MR modes within the single particle gap, for the sequence 1,2,-2,4,-4,5,6  \cite{cooper}. If the corresponding virtual transition lines in a given subphase are reasonably close to the true 
critical line, I find, for example in phase $-4$, {\bf four} modes at $q_{\vert \vert}=2, 7, 8, 9 $. Applying the method of this paper, I find that the mode at $q_{\vert \vert}=2$ should have a much weaker H dependence than the modes 7,8,9.

The occurence of the MR modes of the UQC could have measurable consequences on various properties, such as transport (T dependence of the longitudinal resistivity $\rho_{xx}$), specific heat, NMR relaxation time,) etc.\cite{pl96} \cite{pl97}. The success of the QNM in explaining the phase
diagram and the Hall quantization gives  added confidence that the new type of MR particles described here exists, so that  a renewed experimental effort is called for to detect  them and check theoretical predictions on this unique electron-hole condensate, the Ultra Quantum Crystal. 
\acknowledgments
I am grateful to C. M. Chaves and the Departamento de Fisica da PUC-Rio, to Gilson M. Carneiro and the Instituto de Fisica da UFRJ for their hospitality during the completion of this work.

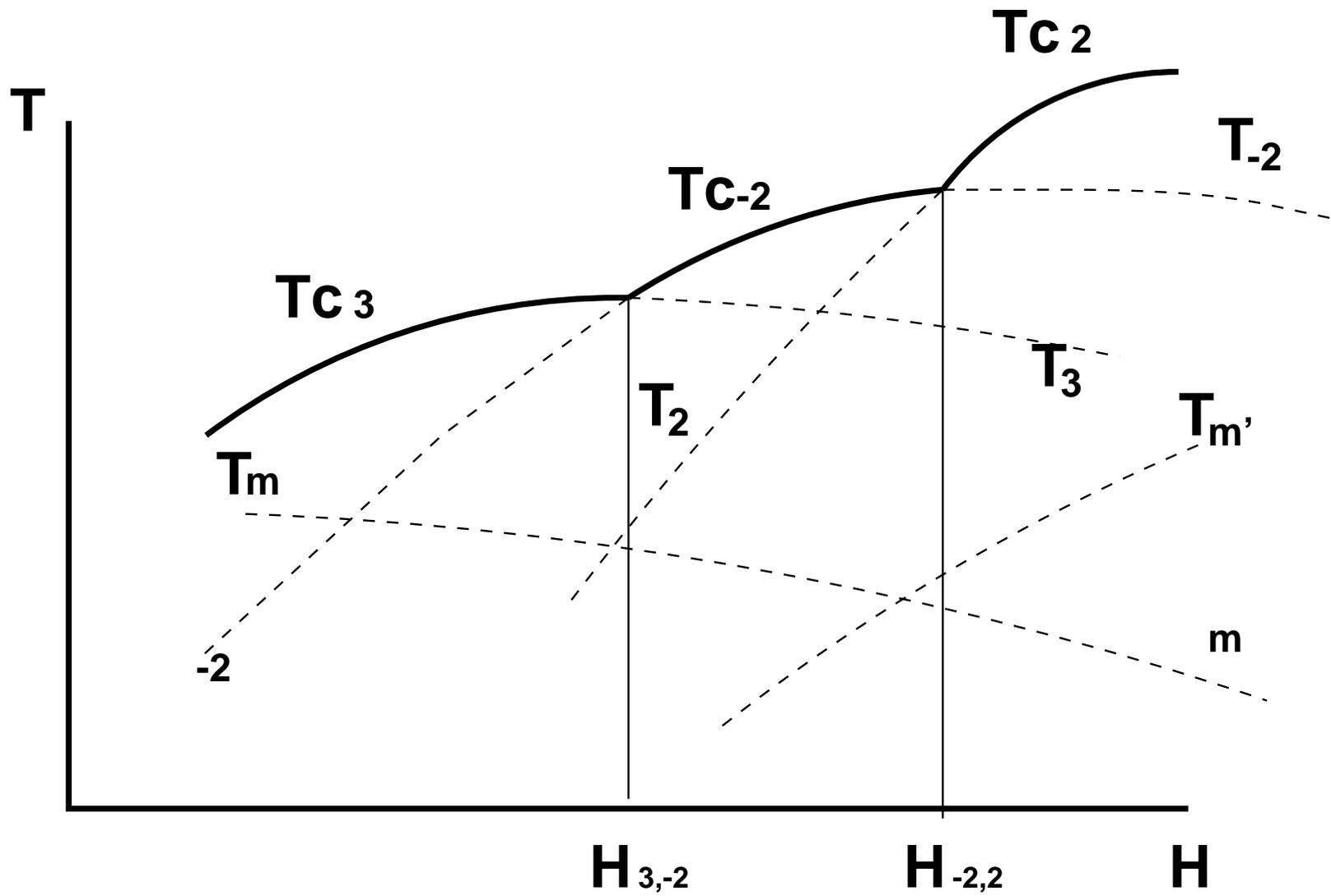
\begin{figure}
\caption{Schematic portion of the (T,H) plane containing the Ribault phase $N=-2$;  actual transition lines (e. g. $T_{c-2}$ ) are full, virtual ones  (e. g. $T_2$)  dotted.  Only two low temperature generic lines are shown, labelled with $m$ and $m'$.
 $H_{3,-2}, H_{-2,2}$ are the transition fields. The lines $T_6$ and $T_7$ (see discussion of Eq.(2.6)), too close to $T=0$, are not shown.}
\end{figure}
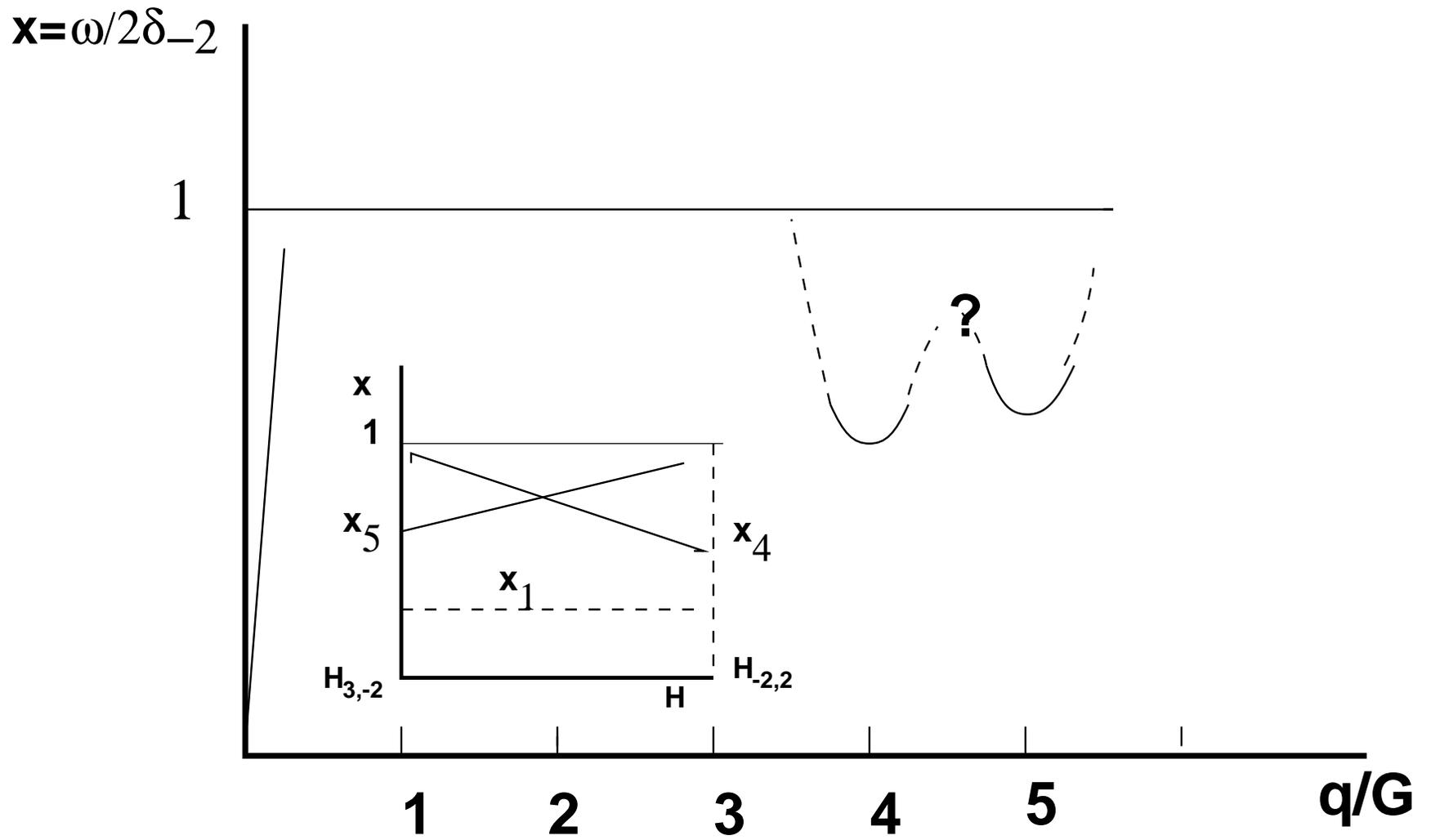
\begin{figure}
\caption{Structure of MR modes in the Ribault phase $N=-2$. The dispersion relation along $q_{\vert \vert}$, within the RPA, is $\omega^2 (\delta) - \omega^2(0) \simeq v_F^2\delta^2 $ around each minimum; 
the structure around  $\delta = G/2$ is not yet known. The inset shows qualitatively the H dependence of the Ribault phase MR modes relative to the single particle gap (continuous lines with slope a few tenths of K per T), as opposed to the usual $q_{\vert \vert}=1$ mode (dotted line)}
\end{figure}


\begin{references}
\bibitem{review}For recent reviews, see L. P. Gork'ov, J.Phys. I France {\bf 6}, 1697 (1996); P. M. Chaikin, {\it ibid}, p. 1875; P. Lederer,{\it ibid}, p. 1899; V. M. Yakovenko and H.S. Goan, {\it ibid}, p. 1917
\bibitem{gorkov} L. P. Gor'kov and A. G. Lebed, J. Physique. Lett. 
{\bf 45}, L433, (1984); P. M. Chaikin, Phys. Rev. {\bf B 31}, 4770 (1985)
\bibitem{gm91}G. Montambaux, Physica Scripta, {\bf T 35}, 188 (1991).
\bibitem{pl96} P. Lederer, J. Phys. I France {\bf 6}, 1899 (1996)
 \bibitem{hlm1}M. H\'eritier, G. Montambaux and P. Lederer, Jour Physique 
 Lett. {\bf 45}, L943, (1984). See also K. Yamaji, J. Phys. Soc. Jap. 
 {\bf 54}, 1034, (1985) 
\bibitem{zm}D. Zanchi and G. Montambaux, Phys. Rev. Lett. {\bf 77}, 366 (1966)
\bibitem{rib2}M. Ribault, Mol. Cryst. Liq. Cryst. {\bf 119}, 91 (1985).
\bibitem{pivetau}B. Piveteau {\it et al} J. Phys. C {\bf 19}, 4483 (1986)
\bibitem{cooper}J. R. Cooper, {\it et al} Phys. Rev. Lett. {\bf 63}, 1984 (1989); S. T. Hannahs 
{\it et al}, Phys. Rev. Lett. {\bf 63}, 1988 (1989).
\bibitem{bali}L. Balicas, G. Kriza and F. I. B. Williams,
Phys. Rev. Lett. {\bf 75}, 2000 (1995).

\bibitem{pl87}P. Lederer and D. Poilblanc, C. R. Acad. Sc. Paris, 
{\bf 304}, II-251 (1987).
\bibitem{dppl} D. Poilblanc and P. Lederer Phys. Rev. 
{\bf B 37}, 9650 (1988); {\bf  B 37}, 9672 (1988)
\bibitem{lm} P. Lederer and G. Montambaux Phys. Rev. {\bf B 37},5375 (1988) 
\bibitem{pl97}A more detailed account will appear in P.Lederer, (in preparation)
\bibitem{plcmc}P.Lederer and C.M.Chaves, (in preparation)

\bibitem{lra}P. A. Lee, T. M. Rice and P. W. Anderson, Sol. St. Com.  
{\bf 14},  703  (1986).


 \end{references}
\end{document}